\documentstyle[aps,psfig,bbox]{revtex}
\begin{document}

\newcommand{\half}{{\textstyle\frac{1}{2}}} 
\newcommand{\gton}{\mathrel{\lower.5ex \hbox{$\stackrel{> }
 {\scriptstyle \sim}$}}}
\newcommand{\lton}{\mathrel{\lower.5ex \hbox{$\stackrel{< }
 {\scriptstyle \sim}$}}}
\newcommand{\ee}{\end{equation}}
\newcommand{\ben}{\begin{enumerate}} \newcommand{\een}{\end{enumerate}}
\newcommand{\bit}{\begin{itemize}} \newcommand{\eit}{\end{itemize}}
\newcommand{\bc}{\begin{center}} \newcommand{\ec}{\end{center}}
\newcommand{\bea}{\begin{eqnarray}} \newcommand{\eea}{\end{eqnarray}}
\newcommand{\beqar}{\begin{eqnarray}} \newcommand{\eeqar}[1]{\label{#1}
\end{eqnarray}} 
\newcommand{\bra}[1]{\langle {#1}|}

\title{Jet Tomography of Au+Au Reactions Including Multi-gluon Fluctuations}

\author{M.~Gyulassy$^{1}$, P.~Levai$^{2}$, and I.~Vitev$^{1}$}

\address{
$^{1}$ Department of Physics, Columbia University, 
       538 W. 120-th Street, New York, NY 10027, USA \\
$^{2}$ KFKI Research Institute for Particle and Nuclear Physics, PO Box 49,
      Budapest 1525, Hungary
}

\maketitle

\begin{abstract}
  Jet tomography is the analysis of the attenuation pattern of high
  transverse momentum hadrons to determine certain line integral
  transforms of the density profile of the QCD matter produced in
  ultra-relativistic nuclear collisions.  In this letter, we calculate
  the distortion of jet tomography due to multi-gluon fluctuations
  within the GLV radiative energy loss formalism.  We find that
  fluctuations of the average gluon number, $\left\langle N^g
  \right\rangle \sim 3$ for RHIC initial conditions, reduce the
  attenuation of pions by approximately a factor $Z\approx 0.4-0.5$.
  Therefore the plasma density inferred from jet tomography without
  fluctuations must be enhanced by a factor $1/Z\sim 2$.

\vspace{.2cm}


\end{abstract}

\section{Introduction}
The discovery of a factor of $\sim 3$ suppression of moderate $p_{\rm T} 
\lton 4$~GeV $\pi^0$'s in central $Au+Au$ reactions by 
PHENIX~\cite{Adcox:2001jp}  and large transverse
asymmetries in non-central collisions for $p_{\rm T} \lton 5$~GeV 
by STAR~\cite{starv2} have inspired
several attempts~\cite{gvw,Levai:2001dc,Fai:2001vz,gvw2,Vitev:2001zn} to begin jet
tomographic analysis of the matter density produced 
in ultra-relativistic nuclear reactions.  Jet tomography is the QCD
analog of conventional X-ray or positron tomography in that it
exploits the attenuation of high energy jets produced in a
variable density medium~\cite{Gyulassy:2001zv}.  The main source of
attenuation of jets in QCD is induced gluon radiation due to multiple
interactions in the medium. While the present data at moderate $p_{\rm T}$
are not conclusive, the tomographic analysis suggests that densities
up to 100 times nuclear densities may have already been achieved. Soon 
very high statistics data out to $p_{\rm T} \sim 10-20$~GeV will be 
obtained, and it is important to refine the theory of jet tomography 
to take into account many sources of distortion of the attenuation 
pattern. In this letter we provide details of the calculation 
of the distortions due to gluon number fluctuations and
compute the correction factors to the deduced densities. 
Such density renormalization factors $Z\sim 0.4-0.5$  
have already been applied to calculations of particle 
spectra at RHIC~\cite{Levai:2000he,conf}.

Jet quenching  probes the   
gluon density of the medium, characterized through the opacity
parameter $L/\lambda_g = \int \sigma_g(\tau) \rho_g(\tau) \, d\tau$. 
For the moderate opacities expected
in nuclear collisions in the $\sqrt{s} \sim 200$~AGeV range, it is
convenient to calculate the induced radiation as a power series
expansion in the opacity.  In Ref.~\cite{glv2b} we derived an analytic
expression for the medium induced gluon radiation spectrum,
$\rho(x)=dN_g/dx=\sum_{n=1}^\infty \rho^{(n)}(x)$, in such a series
form, where $x$ is the light cone 
momentum fraction carried by the radiated gluon (see also~\cite{Wiedemann:2000za} 
and~\cite{Zakharov:2000iz} for the  relation of this approach to the 
asymptotic approach of Ref.~\cite{Baier:2001yt}).  

The result for the $n^{th}$ order in opacity contribution to
$\rho(x)$ for a jet of energy $E$ and color Casimir $C_R$ that is 
produced inside a medium with opacity $L/\lambda_g$
can be expressed from~\cite{glv2b}  in  the following form
\beqar
\rho^{(n)}(x,E) 
&=& \frac{C_R \alpha_s}{\pi^2} \frac{x_c}{x^2} \frac{1}{(n+1)!} 
\left( \frac{L}{\lambda_g} \right)^{n} 
\theta \left(x-\frac{\mu}{2 E}\right)
\int_{K_{\min}}^{K_{\max}} \frac{d^2{\bf k}}{\mu^2} 
 \int \prod_{i=1}^n \left[d^2{\bf q}_{i}\, \frac{d^2\sigma^{eff}({\bf q}_{i})}
{d^2{\bf q}_i} \right]
\, \times \sum_{m=1}^n A_{n,m} \, F_{n,m} \;\;.
\eeqar{ndifdis}
The distribution of transverse momentum impulses
is given by an effective dipole-like differential cross section that 
has an elastic component assumed to be given by a color screened Yukawa 
form~\cite{Gyulassy:1993hr} and a $\delta$-function component in the 
forward (jet) direction  
$\frac{d\sigma^{eff}({\bf q}_{i})}{d^2 {\bf q}_i}
=\frac{\mu^2}{\pi}\frac{1}{({\bf q}_i^2+\mu^2)^2} -
\delta^2({\bf q}_{i})$.
We take $\mu \simeq gT \sim 0.5$~GeV for RHIC initial conditions.

One important advantage of the opacity formalism is that
the effects of a finite temperature QCD plasmon frequency 
cutoff $\omega_{pl}\sim gT/\sqrt{3}$ of the 
gluon radiation can be taken into account approximately 
via the kinematical cut-off both at small $x \lton  x_0 = \frac{\mu}{2 E}$ 
and at small $k \lton \mu=K_{\min}$. We ignore the $\sim \sqrt{3}$ 
difference between the plasma frequency and the screening mass since 
$g \sim  2$ and approximate both by $\sim 0.5$~GeV. 
The upper bound $K^2_{\max}(x)=4E^2\min(x^2,(1-x)^2)-\mu^2$
in $k^2$ results from requiring the quenched jet as well as the radiated 
gluons have positive forward momenta. The 
fraction $x_c/x\equiv \mu^2 L/(2xE)$ is a measure of the thickness
of the medium to the gluon formation length.    

The radiation amplitudes are 
here denoted by
$A_{n,m}= 2\,({\bf k}-{\bf Q}_{n}) \cdot({\bf C}_{n,m+1}-{\bf C}_{n,m})$,
where ${\bf Q}_{n}\equiv {\bf q}_{1}+\cdots+{\bf q}_n$, with 
${\bf Q}_{0}\equiv 0$, and 
${\bf C}_{n,m}= \half \nabla_{{\bf k}} \log ({\bf k} - {\bf Q}_{n}+{\bf Q}_{m-1})^2$.
Destructive interference suppresses radiation with
formation times greater than the thickness, $L$, 
of the medium. In addition to one power of $x_c/x$ in Eq.(1),
higher order contributions are further suppressed by a formation factor
from Eq.~(116) of~\cite{glv2b}
\beqar
F_{n,m} \equiv {\rm Im} \; \prod_{j=1}^m 
\left(1+ i\, \frac{x_c}{x} 
\frac{({\bf k}-{\bf Q}_n+{\bf Q}_{j-1})^2}{ \mu^2 (n+1) } 
\right)^{-1}  \;\;.
\eeqar{fm}
This simple analytic form arises  for an  exponential  
distribution, $\propto \exp\left(-(z_{k}-z_{k-1})\frac{L}{n+1}\right)$, 
between adjacent scattering centers, $z_k$,
in a plasma with {\em mean} thickness $L$ at $n^{th}$ order in opacity.

Gluon reabsorption from the
medium reduces the radiation density $\rho(x)$ at low $x$
as shown in~\cite{Wang:2001cs}. However, this effect is only important for jets
with $E$ less than a few GeV. Here we focus on higher energy jets.
We consider only the {\em static} plasma geometry in Eq.~(2) to simplify the
numerical evaluation of $\rho^{(n)}$ at higher orders. In
Refs.~\cite{gvw,gvw2,Gyulassy:2001zv} we showed that the the main
effect of 3+1D (Bjorken+transverse) expansion in the opacity expansion
is to reduce the mean radiative energy loss, $\Delta E = Z_{3+1} \cdot \Delta
E_{stat.}$ relative to the static approximation by a renormalization
factor $Z_{3+1}= 2\tau_0/L$, where $\tau_0$ is the formation time of
the matter. The {\em effective} static opacity $L/\lambda=5$ that we
use to fit the PHENIX data is relatively small because it value reflects
the rapid dilution effects due to expansion.

The numerically computed  mean number  of radiated gluons and 
the mean energy  loss up to third order,
\beqar
\left[ \begin{array}{c} \Delta E  \\ 
    \left\langle N^g \right\rangle   \end{array} \right] =\int dx
\left[ \begin{array}{c} x E 
\\ 1  \end{array} \right]   \left( \rho^{(1)}(x,E)
+\rho^{(2)}(x,E)+\rho^{(3)}(x,E)\right) \;\;.
\eeqar{defe}
are shown in Figs.~1 and 2. 
The curves are calculated using $\mu=0.5$~GeV, 
$\lambda_g=1$~fm, and $C_R=3$ and running $\alpha_s$.
The leading first order gluon energy loss $\Delta E^{(1)}$, 
is found to 
roughly follow the leading log expression
$\Delta E^{(1)}/ E \approx 
\frac{3\alpha}{4}\frac{\mu^2 L^2}{\lambda_g}
\log\frac{1}{x_c}$ from~\cite{gvw2,Zakharov:2000iz}.
The second order contribution reduces the first order result by a factor
$\sim 2$ for $E\lton 10$ GeV. By 40 GeV the second order correction 
is only 10\%. For $E > 40$ both the second and third order 
correction are negligible, but also for $E\sim 5-20$ GeV the third 
order contribution largely cancels the second order one. 
The summed first through third orders shows that the induced fractional 
energy loss varies from $0.5$ at $E \sim 5$~GeV down to 
$\sim 0.3$ at $20$~GeV. Jet tomography at RHIC is dominated by quark 
fragmentation for pions with  $p_{\rm T} > 5$~GeV with $\Delta E$ 
reduced by a factor $C_F/C_A = 4/9$.
\vspace{-1.2cm}
\begin{center}
\includegraphics{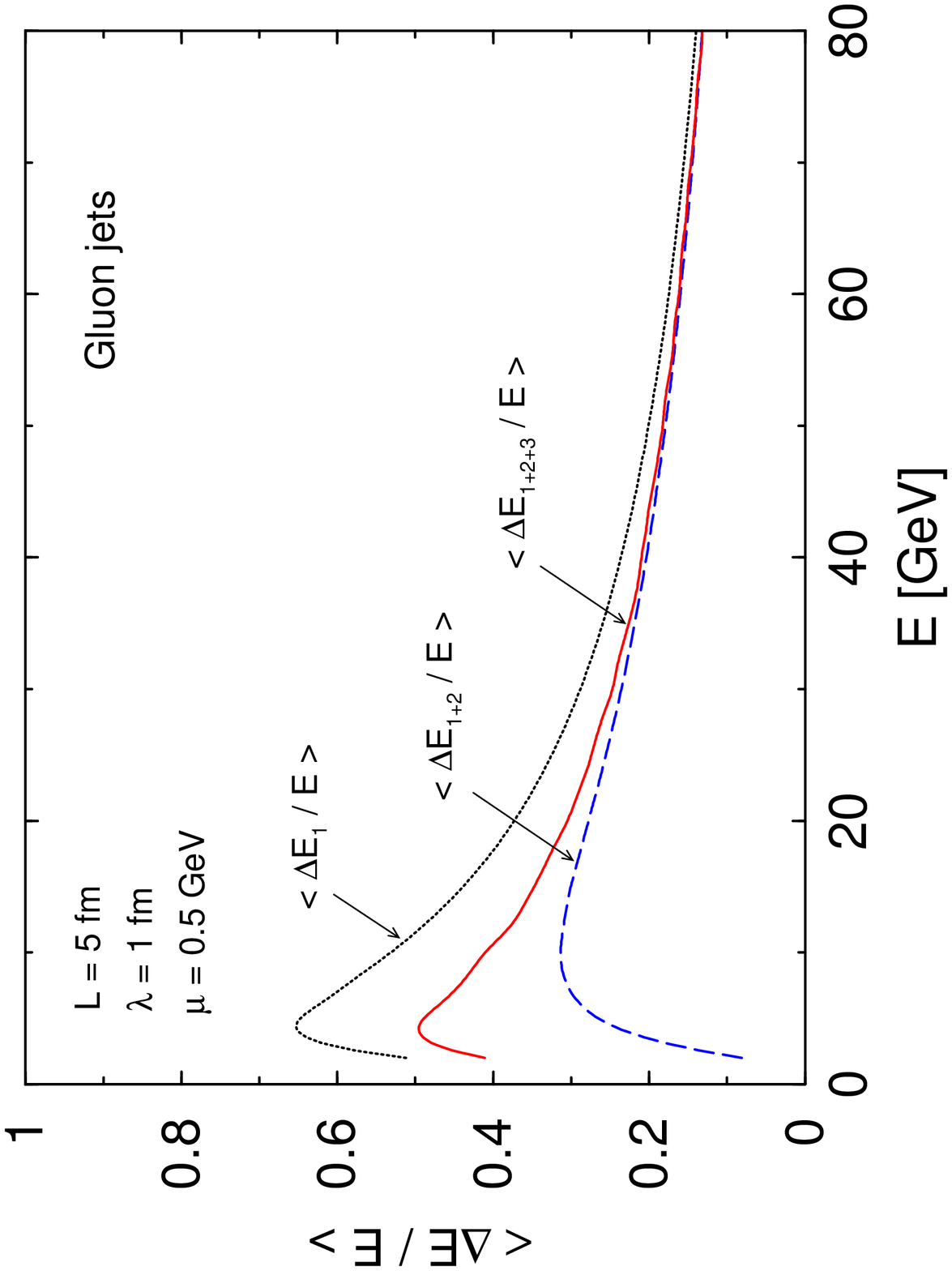}
\vspace{9.8cm}
\end{center}

\begin{center}
\begin{minipage}[t]{12cm}
         { FIG. 1.} 
         {\small $\Delta E/E$ is plotted versus  $E$ for opacity  
$L/\lambda_g=5$. The three curves correspond to calculations up to
1 order (upper bound), 1+2 order (lower bound), and  
1+2+3 order (the actual final result). }
\end{minipage}
\end{center}

A similar pattern is seen in the energy dependence of the average number 
of gluons radiated in Fig.~2. The second and third order corrections largely
cancel and the final average gluon number is only 2-3 for the kinematic 
range accessible at RHIC.  Even with sensitivity $\sim 2$ on 
the kinematic cut-offs,
$\left\langle N^g \right\rangle$ is small.  In fact as a function of 
energy the gluon number saturates for any given opacity $L/\lambda_g$.  
The gluon radiative distributions 
are strongly  peaked at small $x$ (which is consistent with the small $x$ 
approximations used) and naturally need a lower cut-off generated by
a characteristic jet energy independent mass 
scale $m$, i.e. $x_{\min} \sim m/E$.   
For any gluon radiative distribution that has a form 
$\rho(x,E)\approx 1/x_c f(x/x_c)$ 
(with $x_c={\rm typical\; energy \;  scale/E}$)  
the mean number of  gluons due to induced
radiation $\left\langle N^g \right\rangle$ is approximately 
jet energy independent.   
As the energy of the jet and the radiative energy loss increase  
$\left\langle N^g \right\rangle$ remains small
but the radiated  gluons become harder. 

We note that  both $\Delta E$ and 
$\left\langle N^g \right\rangle$ are much smaller than 
the recent estimates in~\cite{Baier:2001yt} 
because our {\em effective} 
transport coefficient $\mu^2/\lambda_g\approx 0.25 \; {\rm GeV}^2$/fm, 
which is constrained by our 
fit below (see Fig.~4) to the PHENIX pion attenuation data~\cite{Adcox:2001jp},
is 4 times smaller than the one considered for illustration 
in~\cite{Baier:2001yt}.
As noted before, expansion greatly reduces
estimates based on the transport properties of the high density 
initial conditions. In addition, the finite plasmon frequency cut-off,
$\sim \mu$, is the  medium  regulator of  soft gluon number divergences. 
We emphasize that it is
the finite medium dependent cut-off of the soft spectrum together
with the relatively small transport coefficient and opacity of the plasma
produced at RHIC energies that allows 
the opacity expansion to converge so rapidly 
even under the extreme conditions produced at
RHIC.

\vspace{-1.2cm}
\begin{center}
\includegraphics{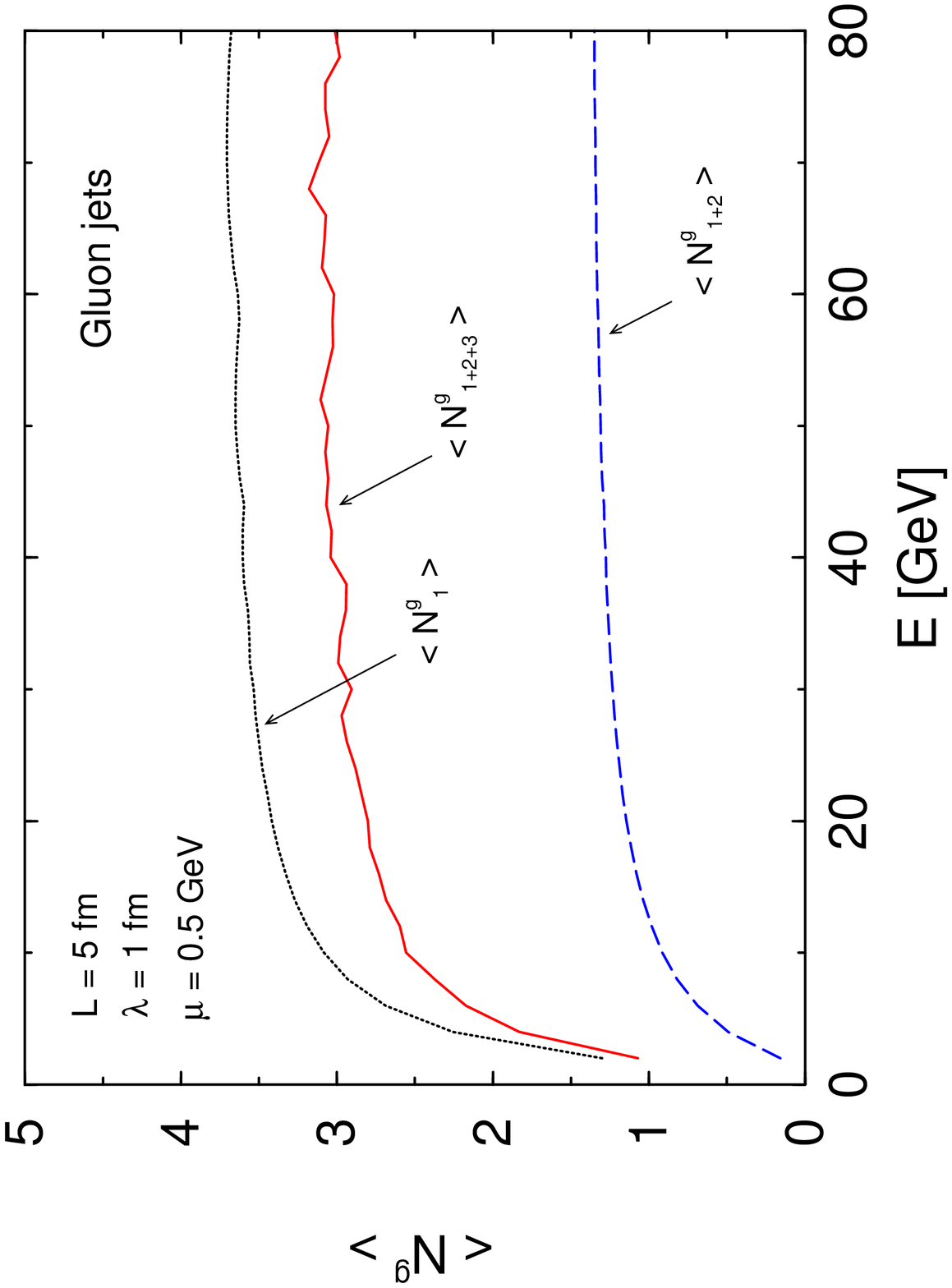}
\vspace*{9.8cm}
\end{center}

\begin{center}
\begin{minipage}[t]{12cm}
         { FIG. 2.} 
         {\small Average number of gluons, $\langle N^g(E)\rangle$,
 with $\omega \gton \mu$ is plotted versus  E for opacity  $L/\lambda_g=5$. 
A set of curves corresponding to the curves in Fig.~1 is presented.}
\end{minipage}
\end{center}

\section{Fluctuation spectrum of radiative  energy loss}

In the approximation
 that the fluctuations of the gluon number are 
uncorrelated, the spectrum of the total radiative energy loss fraction, 
$\epsilon=\sum_i \omega_i / E$,
can be expressed via a  Poisson expansion
$P(\epsilon,E)=\sum_{n=0}^\infty P_n(\epsilon,E)$ 
with $P_1(\epsilon,E)=e^{-\left\langle N^g \right\rangle} \rho(\epsilon,E)$ and
\beqar
P_{n+1}(\epsilon,E)&=&  \frac{1}{n+1} \int_{x_0}^{1-x_0} dx_n \; \rho(x_n,E)
P_n(\epsilon-x_n,E) 
\nonumber \\[1ex]
&=&\frac{e^{-\left\langle{N^g(E)}\right\rangle}}{(n+1)!}\int dx_1\cdots 
dx_{n} \; \rho(x_1,E)\cdots\rho(x_{n},E)\rho(\epsilon-x_1-\cdots-x_{n},E)
 \;\;.
\eeqar{pep}
The form of this spectrum guarantees that the mean value
is as in Fig.~1:
\beqar 
\int_0^\infty d\epsilon \; P(\epsilon,E) \epsilon= \frac{\Delta E}{E}
\;\; .
\eeqar{sumpe}
The above distribution differs considerably from that computed 
in~\cite{Baier:2001yt} because $\langle N^g(E) \rangle$ is finite and small 
in our case and because $\rho(x)=0$ for $x<\mu/(2E)$ 
due to the plasma frequency cutoff that we impose on the 
low frequency modes in the medium. 
Therefore, we have explicitly a finite $n=0$ (no radiation) 
contribution $P_0(\epsilon,E)=
e^{-\left\langle {N^g(E)}\right\rangle }\delta(\epsilon)$.
Assuming negligible kinematic correlations the
numerical iteration of the recursion relation in Eq.~(4)  becomes very fast
and is computed to  high order ($n  \leq 25$).
However, by not enforcing that $\sum_i x_i \leq 1$,
there is  a ``leakage'' error into the unphysical  $\epsilon>1$ range.
We calculate this ``leakage'' error $\int_1^\infty d \epsilon \;P(\epsilon,E)$
 and correct the normalization of $P(\epsilon,E)$ in the 
the physical range $\epsilon \in [0,1]$.

The resulting spectrum (without the delta function contribution) 
is shown in Fig.~3 for three different jet energies. The finite intercept 
at $\epsilon = 1$ provides a measure of the ``leakage'' error, 
which is acceptable in this case. The low frequency plasmon 
cut-off at $ \sim \mu$ is  clearly visible. In addition 
multi-gluon iterations in the probability distribution $P(\epsilon,E)$
exhibit a slight oscillatory pattern in multiples of $x_0$. 
The high frequency  random oscillations provide an indication 
of the accuracy of our Monte Carlo numerical integrations 
methods.  

The results indicate that $P(\epsilon,E)$ is approximately constant from 
$x_0=\mu/(2E)$ up to a scale $ \sim x_c = \mu^2L/(2E)$. 
For $x \gg  x_c$, $P(\epsilon,E)$ decreases rather
quickly. The (normalized to unity) probability distribution per gluon, 
$\rho(x,E=40\;{\rm GeV})/ \left\langle N^g \right\rangle $ 
is also shown for comparison. Multi-gluon fluctuations flatten
the rapid small $x$ rise of $\rho$ even though 
Eqs.~(\ref{defe},\ref{sumpe}) dictate
that the first moment of both 
$\rho(x,E)$ and $P(\epsilon,E)$ are the same. 
\begin{center}
\includegraphics{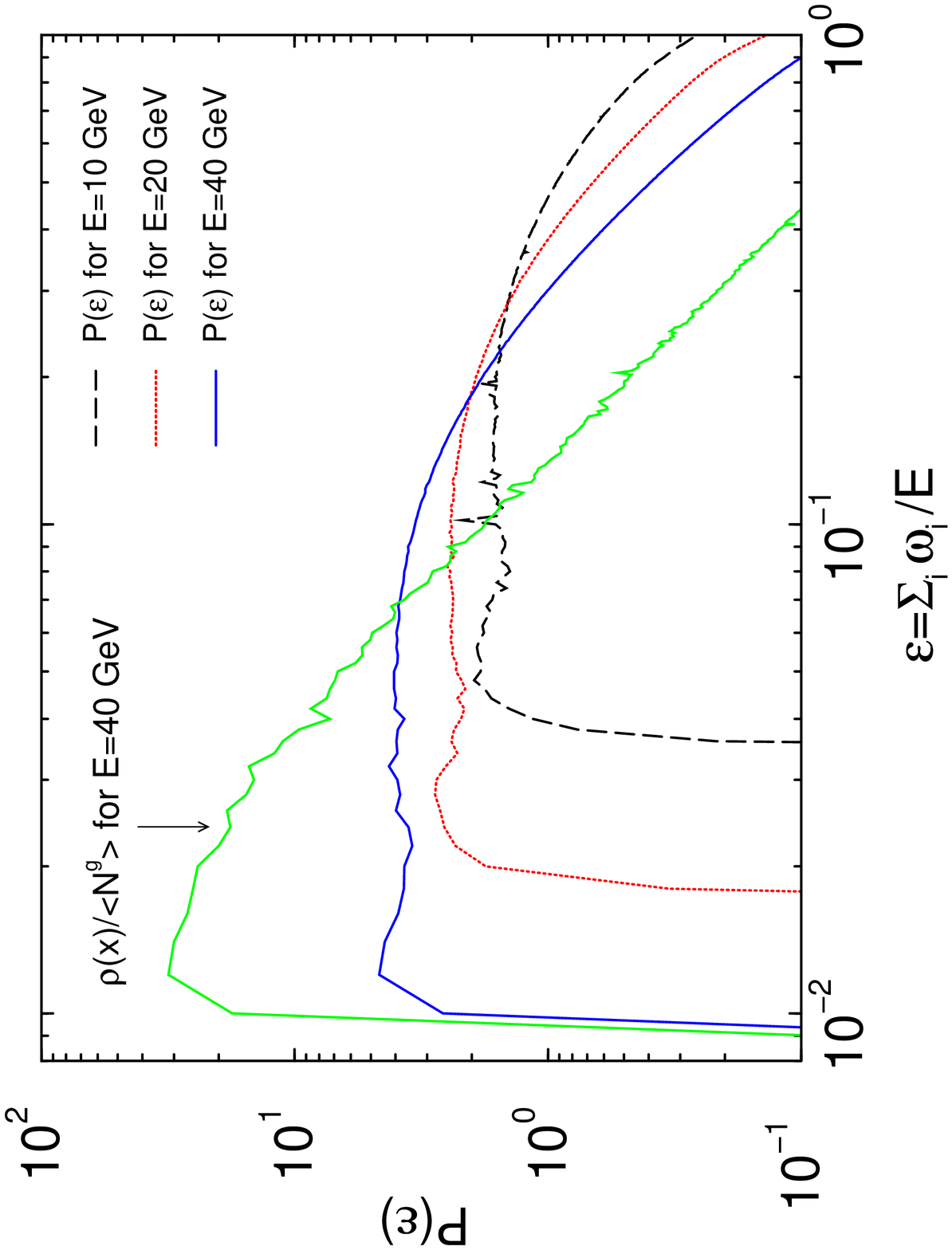}
\vspace{8.6cm}
\end{center}

\begin{center}
\begin{minipage}[t]{12cm}
         { FIG. 3.} 
         {\small Probability density of total fractional energy loss 
$\epsilon=\sum \omega_i/E$ for a gluon get with E=10,20, and 40 GeV 
traversing matter 
with  opacity  $L/\lambda_g=5$. The numerical curves include orders 1+2+3 order
in the opacity expansion for $\rho(x)$. The $n=0$ no radiation delta function
contribution, $P_0(\epsilon)$, is not shown above. The low frequency plasma
cut-off is at $\epsilon=\mu/(2E)$. The probability density per gluon is also
included.}
\end{minipage}
\end{center}

\section{The Quenching Pattern of $\pi^0$}

We apply the energy loss spectrum to calculate the
quenched spectrum of hadrons
by modifying the mean energy loss pQCD formulas  from 
Refs.~\cite{gvw,Levai:2001dc,Fai:2001vz,gvw2,Vitev:2001zn,WaHu97}. 
We concentrate on mid-rapidity hadron production ($y_{cm}=0$).
A jet of flavor $c$ and  transverse
momentum $p_c$ produced in a hard PQCD scattering $a+b\rightarrow c+d$
is attenuated prior to hadronization by the radiative 
energy loss to $p_c^* = p_c (1-\epsilon)$. This 
shifts the hadronic fragmentation fraction $z_c=p_h/p_c$ 
to $z_c^* = z_c /(1-\epsilon)$.

The invariant distribution of $\pi^0$ reduced by energy loss
in central $A+A$ collision is then given by
\begin{eqnarray}
\label{fullaa}
&&E_{h}\frac{dN_{\pi^0}^{AA}}{d^3p} = T_{AA}(0)
        \sum_{abcd}\!  
        \int\!\!dx_1 dx_2  \;  
f_{a/A}(x_1,Q^2) f_{b/A}(x_2,Q^2)\
             \frac{d\sigma^{ab \rightarrow cd}}{d{\hat t}} 
\int d\epsilon \; P(\epsilon,p_c)
\frac{z^*_c}{z_c}
   \frac{D_{\pi^0/c}(z^*_c,Q^2)}{\pi z_c} \,\,\, ,
\end{eqnarray}
where $T_{AA}(0)$ is the Glauber profile density in
central collisions. 
The pion fragmentation function $D_{\pi^0/c}(z,Q^2)$ is taken 
from BKK~\cite{BKK95}.
We take the GRV94 LO~\cite{GRV94} structure functions for $f_{a/p}(x,Q^2)$ 
and include isospin dependence ($Z$ protons and $A-Z$ neutrons).
Nuclear shadowing, intrinsic $k_{\rm T}$ broadening and Cronin effect can be
taken into 
account as  in~\cite{XNWint,PLF00,Gerg00,Eloke}.  
The interplay between  the soft and
hard components of hadron production studied in~\cite{Vitev:2001zn,conf} 
lead to modifications of the spectral shapes  
in the  low $p_{\rm T}$ region and are neglected in this analysis.
The factor $z^*_c/z_c$ appears because of  the in-medium 
modification of the fragmentation function~\cite{WaHu97}.
Thus, the invariant cross section Eq.~(\ref{fullaa})  depends on the average
opacity $L/\lambda_g$ through the effect of $P(\epsilon,p_c)$.

We consider three different approximations to $P(\epsilon,E)$:
\begin{enumerate}
\item Use only the mean energy loss with $P(\epsilon,E) \approx
\delta(\epsilon-\Delta E(E) / E) $ as in  
\cite{gvw,Levai:2001dc,Fai:2001vz,gvw2,Vitev:2001zn,WaHu97}
\item Use the full fluctuating spectrum, $P(\epsilon,E)$,  from Eq.~(\ref{pep}) 
\item Use a renormalized average energy loss with 
$P(\epsilon,E,Z) \approx
\delta(\epsilon - Z \cdot \Delta E(E)/E)$
\end{enumerate}

\begin{center}
\vspace*{0.0cm}
\includegraphics{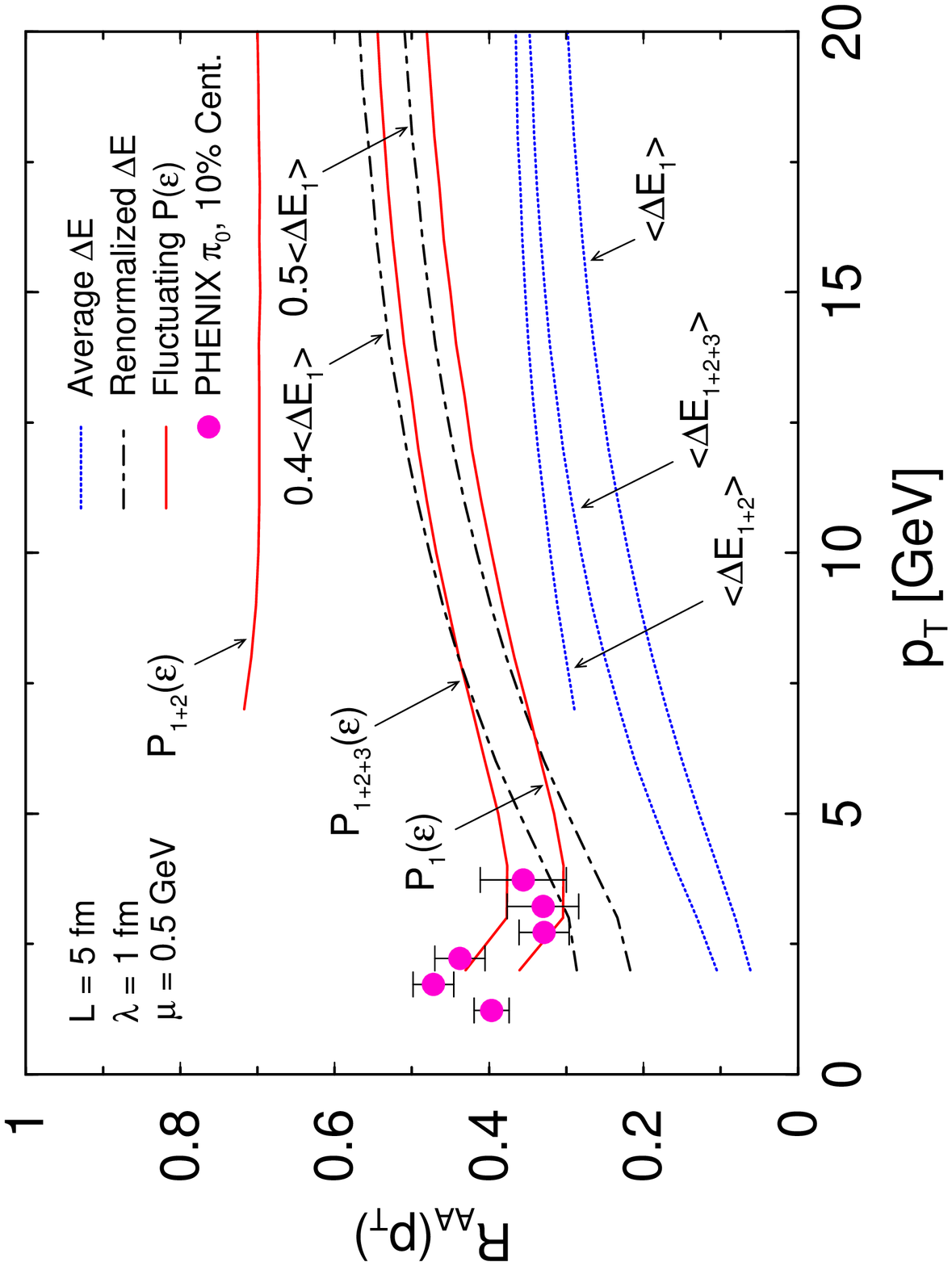}
\vspace{8.8cm}
\end{center}

\begin{center}
\begin{minipage}[t]{12cm}
         { FIG. 4.} 
         {\small Quenching pattern for $\pi^0$ versus transverse momentum order
by order in the opacity expansion. Curves labeled $<\Delta E_{1+\cdots}>$
are calculated using the average energy loss up to the order indicated.
The average opacity  is taken to be $L/\lambda_g=5$. The data are from 
PHENIX~\protect{\cite{Adcox:2001jp}}. Curves labeled $P_{1+\cdots}(\epsilon)$ 
average over the fluctuating distributions illustrated in Fig.~3. The dot-dashed curves
correspond to using a reduced (Z renormalized) first order average energy loss. }
\end{minipage}
\end{center}

The ratio, $R_{AA}(p_{\rm T})$, compares the quenched to the unquenched
$\pi^0$ distributions. In the case (1), the convergence of the opacity
series using the mean energy shift to first and up to third order
appears to be reasonably fast even though the second order correction
is still uncomfortably large below $p_{\rm T} \lton 10$ GeV.  Improved
numerical methods need to be developed to enable summing higher order
terms to verify our expectation that the summed results to third order
are not significantly changed by higher order due to the additional
$1/(n+1)$ and $F_{n,m}$ factors in Eq.~(1).
 
We see from Fig.~4 that with even the modest value of the opacity 
$L/\lambda_g=5$, the mean energy loss approximation over 
predicts the observed quenching by about a factor of two.
Including the fluctuations in the Poisson approximation 
via $P(\epsilon,E)$ leads to less energy loss by approximately a factor 
of two and brings the attenuation in line with the observed results.
This renormalization of the effective energy loss can be inferred from 
the dot-dashed curves using approximation (3) above with 
$Z \approx 0.4-0.5$. We conclude that the distortion of jet tomography
due to gluon number fluctuations in the Poisson approximation 
can be well approximated by renormalizing the mean energy loss calculations
by a factor
$Z \sim 0.5$.

While the $p_{\rm T}$ range of the available data is still too low to
draw definitive conclusions, the effective static opacity 
with gluon fluctuation renormalization above corresponds
from the results of ~\cite{Vitev:2001zn} to an estimated
initial gluon rapidity density 
$dN^g/dy\sim 800\pm 100$ and implies that the initial gluon density 
produced at RHIC may have reached 
$\rho_g\approx (dN^g/dy)/(\tau_0\, 
\pi R^2)\sim 20/{\rm fm }^3 \sim 100 \rho_A$.

\acknowledgments

This work was supported by the Director, Office of Science, 
Office of High Energy and Nuclear Physics,
Division of Nuclear Physics, of the U.S. Department of Energy
under Contract No. DE-FG02-93ER40764 and by the U.S. NSF under INT-0000211 
and  OTKA No. T032796.

\vfill\eject
\end{document}